\documentstyle[aps,prl,multicol,epsfig]{revtex}
\begin{document}
\title{Operating a phase-locked loop controlling a high-Q tuning fork
sensor\\ for scanning force microscopy}
\author{T. Ihn, T. Van\v{c}ura, A. Baumgartner, P. Studerus, K. Ensslin}
\address{Laboratory of Solid State Physics, ETH H\"onggerberg, 
CH-8093 Z\"urich, Switzerland}
\maketitle
\begin{abstract}
The implementation of a tuning fork sensor in a scanning force 
microscope operational at 300 mK is described and the harmonic 
oscillator model of the sensor is motivated. These sensors exhibit 
very high quality factors at low temperatures. The nested feedback 
comprising the sensor, a phase locked loop and a conventional 
$z$-feedback is analyzed in terms of linear control theory and the 
dominant noise source of the system is identified. It is shown that 
the nested feedback has a low pass response and that the optimum 
feedback parameters for the phase-locked loop and the $z$-feedback can 
be determined from the knowledge of the tuning fork resonance alone 
regardless of the tip shape. The advantages of this system compared 
to pure phase control are discussed.
\end{abstract}

\begin{multicols}{2}
\narrowtext
\section{introduction}
Since its invention by Binnig, Quate and Gerber \cite{binnig}, the 
scanning force microscope (SFM) has become a standard tool for the 
investigation of conducting and insulating surfaces on the atomic 
scale. Various imaging modes exist that can be classified to be either 
contact modes or dynamic modes \cite{albrecht}. The latter are characterized by the 
fact that the tip oscillates above the surface and the parameters of 
the oscillation are influenced by the tip-sample interaction. True 
atomic resolution has been achieved in this imaging mode 
\cite{giessibl}. Usually the cantilever oscillation is measured by 
optical means. Tuning fork sensors offer the 
possibility of non-optical detection of the cantilever oscillation 
via the piezoelectric effect \cite{dransfeld,giessibl2,rychen1}.
It was demonstrated that atomic resolution is possible with these 
unconventional and very stiff sensors \cite{giessibl1}. The 
fundamental limits to force detection with quartz tuning forks were 
disussed by Grober and co-workers in Ref. \cite{grober}.

We employ 
the tuning fork sensors in a cryo-SFM setup in which the home-built microscope is 
operated in a $^3$He-cryostat at a temperature of 300 mK. Tuning fork sensors
can be classified by the strength of the mixing between 
symmetric and antisymmetric mechanical oscillation modes. In the 
extreme case, one of the tuning fork prongs is firmly attached to a 
support \cite{giessibl2} and extremely strong mode mixing occurs.
In the first section of this paper we present evidence, that our sensors are in the 
weak mode mixing limit and that the simple harmonic oscillator model 
is therefore appropriate. In the second section we describe our nested 
SFM feedback system comprising the tuning fork sensor, a phase-locked 
loop and a conventional $z$-feedback. We present an analysis of the 
linear response and the noise of these feedbacks and show how the optimum settings 
for the feedback parameters can be found. These optimum parameters 
are found to be independent of the tip shape and the details of the 
tip sample interaction. The advantages of using a feedback with 
phase-locked loop for the very high $Q$ tuning fork sensors are 
discussed and a comparison is made with the phase control mode.

\section{Tuning fork sensors: a realistic model}
\begin{figure}[htbp]
  \centering
  \epsfig{file=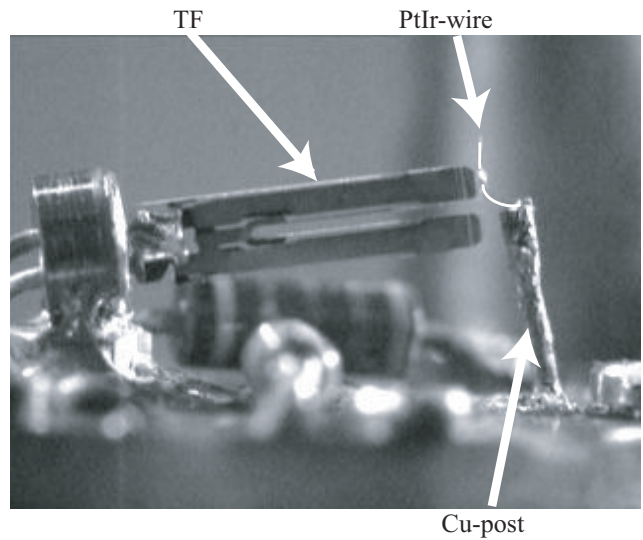,width=3.3in,angle=0}
  \caption{Tuning fork sensor used in our cryo-AFM.}
  \label{fig5}
\end{figure}

In our dynamic mode SFM we employ the same type of
tuning fork sensors previously discussed and calibrated in Refs. 
\cite{rychen1,rychen2}. Figure \ref{fig5} shows a typical sensor. The 
commercially available tuning fork \cite{fork} is mounted on a print plate under an angle of about 
10$^\circ$. A 15 $\mu$m PtIr wire is glued \cite{glue} to the end face of one 
tuning fork prong and to a copper post which serves as the electrical 
contact to the tip. The free end of the wire is then etched 
electrochemically resulting in a sharp tip which will serve as a probe 
for the sample surface. The preparation technique impairs the symmetry
of the tuning fork and we have to care about the effects on the 
oscillation properties.

\begin{figure}[htbp]
  \centering
  \epsfig{file=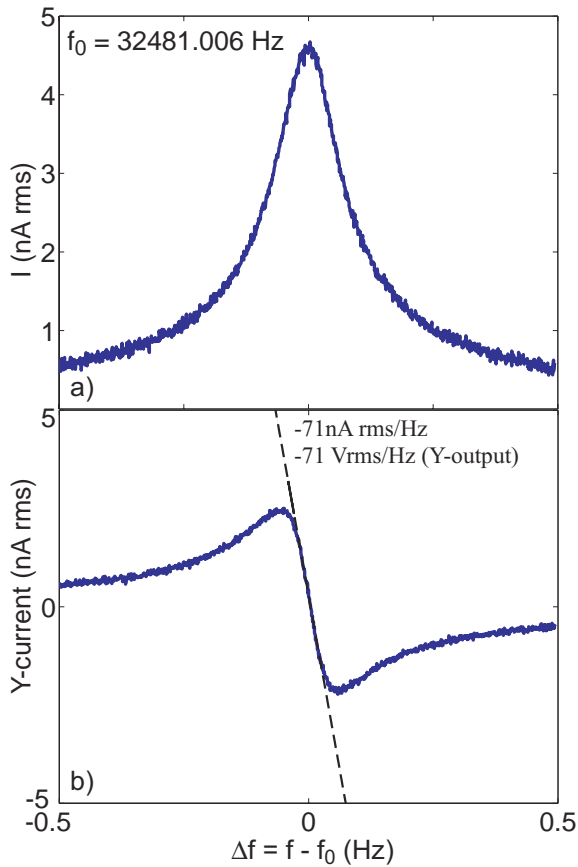,width=3in,angle=0}
  \caption{a) Current magnitude near the tuning fork resonance. 
  b) Y-component of the current near resonance}
  \label{fig67}
\end{figure}

Figure \ref{fig67} shows the current through the tuning fork near 
resonance measured 
at a temperature of 4.2 K with an excitation voltage of $U_{0}=10$ 
$\mu$Vrms applied to the tuning fork contacts. 
The top graph is the magnitude of the current. The experimental details of the 
admittance measurements will be discussed later in the paper. In the following, we 
start with a mechanical model of the tuning fork sensor which eventually 
aims at the detailed understanding of such resonance curves. In 
particular, we will focus on the symmetry breaking effect of the tip 
prepraration.

We describe the mechanical oscillator with a coupled harmonic 
oscillator model driven by the voltage $U_{0}$ via the piezoelectric
coupling constant $\alpha$.
\begin{eqnarray}
    m_{1}\ddot{x}_{1} + \gamma_{1}\dot{x}_{1} + k_{1}x_{1} +k_{c}x_{2} 
     + \gamma_{c}\dot{x}_{2} & = & -\alpha U_{0}\nonumber\\
    m_{2}\ddot{x}_{2} + \gamma_{2}\dot{x}_{2} + k_{2}x_{2} + k_{c}x_{1} 
     + \gamma_{c}\dot{x}_{1} & = & +\alpha U_{0}+F_{ts}(x_{2}).\nonumber
\end{eqnarray}
The deflection of the two prongs is described by the coordinates $x_{i}$.
The $m_{i}$ are the effective masses of the two prongs, the $k_{i}$ 
and $\gamma_{i}$ are their spring- and damping-constants, 
respectively. The tip is attached to prong 2 and interacts with the 
sample surface via the interaction force $F_{ts}(x_{2})$. Coupling 
between the prongs is described by coupling constants $k_{c}$ and 
$\gamma_{c}$.

The Eigenmodes of this oscillator (determined with $F_{ts}(x_{2})=0$) are given by
\[ \Omega_{1,2}^2=\omega_{0}^2 \pm\omega_{c}^2\sqrt{1+\kappa^2}\]
where $\omega_{0}^2=(k_{1}/m_{1}+k_{2}/m_{2})/2$, 
$\omega_{c}^2=k_{c}/\sqrt{m_{1}m_{2}}$ and 
$\kappa=(k_{1}/m_{1}-k_{2}/m_{2})/(2k_{c}/\sqrt{m_{1}m_{2}})$.
In the case of $\kappa=0$, in particular when the two prongs are 
identical, the two Eigenmodes are the symmetric (the $x_{i}$ 
oscillate with zero phase difference) and the antisymmetric (the 
$x_{i}$ oscillate with a phase difference of $\pi$) 
mode. The metallic tuning fork contacts are arranged such that only 
the antisymmetric mode is excited by a driving voltage. For $\kappa\ll 
1$ the two modes do not mix significantly and the symmetries of the 
modes are not strongly affected, while for $\kappa\gg 1$ each prong 
has its own resonance frequency where the other is hardly excited.

In the following we show that our tuning fork sensors are in the 
regime of negligible mode mixing, because the introduced asymmetry is
small, i.e. $\kappa\ll 1$.
We can rewrite the expression for $\kappa$ introducing $m_{1}=m$, 
$m_{2}=m+\Delta m$, $k_{1}=k$, $k_{2}=k+\Delta k$. Expanding up to 
first order in $\Delta m/m$, $\Delta k/k$ and $k_{c}/k$ we obtain
\[ \kappa = \frac{\Delta m/m-\Delta k/k}{2k_{c}/k} \]
It can be seen that the additional mass on one prong can be 
compensated by increasing the spring constant, a fact exploited
already in Ref. \cite{shelimov}.

We estimate the additional spring constant $\Delta k$ of a thin 
wire attached to one prong. It is given by \cite{sarid}
$ \Delta k = 3\pi Er^4/(4L^3)$, 
where $L$ denotes the length of the wire, $E$ is Young's modulus of 
the material and $r$ is its radius. Typical wires have $r=7.5$ $\mu $m, 
$L=0.5$ mm, and $E=200-400$ GPa leading to $\Delta k=12-24$ N/m. 
With $k=14000$ N/m this results in $\Delta k/k\approx 10^{-3}$.

An estimate of the additional mass added to one arm of the tuning 
fork takes essentially the glue into account. The idealized drop of 
glue has the shape of a half-sphere with radius $r$ and volume
$V=2\pi/3r^3$.
Using the mass density of the H20E-epoxy $\rho=2600$ kg/m$^3$
and a radius of 100 $\mu$m we obtain a mass
of 5.44 $\mu$g. This weight has to be related to the mass of a single 
tuning fork arm which is 1.1 mg. This leads to the
ratio $\Delta m/m=5\times 10^{-3}$.

The ratio $k_{c}/k$ can be estimated from finite element simulations 
of tuning fork sensors to be of the order of several percent \cite{teru}.

From these estimates it seems to be reasonable that the employed 
sensors are in the limit of small mode mixing. An additional 
observation strongly supports this point of view: only very rarely is a second 
resonance observed in the admittance. Since the measured current 
detects only the antisymmetric component of the oscillation, strong 
mode mixing would make the second resonance visible.

Motivated by these estimates we 
approximate the equations of motions by $m_{1}=m_{2}=m$, 
$k_{1}=k_{2}=k$ and $\gamma_{1}=\gamma_{2}=\gamma$ and obtain
\begin{eqnarray}
    m\ddot{\xi} + (\gamma-\gamma_{c})\dot{\xi} + (k-k_{c})\xi & = & 
    2\alpha U+F_{ts}(s+\xi/2)\label{eq1}\\
    m\ddot{s} + (\gamma+\gamma_{c})\dot{s} + (k+k_{c})s & = & 
   \frac{1}{2}F_{ts}(s+\xi/2)\nonumber    
\end{eqnarray}
with $s:=(x_{1}+x_{2})/2$ is the centre of mass coordinate and 
$\xi=x_{2}-x_{1}$ is the relative coordinate.
The tip-sample interaction 
force couples the centre of mass motion to the relative motion.
However, due to the extremely high stiffness of our tuning forks (of 
the order of $10^4$ N/m) a typical oscillation amplitude of 1 nm gives 
rise to a restoring force of $10^{-5}$ N. This is very large 
compared to typical tip-sample interaction forces of the 
order of 1 nN. Therefore we neglect the coupling and describe the 
tuning fork oscillation with the approximation $s=0$ which implies 
that $x_{1}=-x_{2}$ and $\xi=2x_{2}$. Inserting these relations into 
equation (\ref{eq1}) leads to the single harmonic oscillator approximation
for our tuning fork sensors, which we rewrite with $x=x_{2}$, 
$(k-k_{c})/m=\omega_{0}^2$ and $(\gamma-\gamma_{c})/m=\omega_{0}/Q$
\[ \ddot{x} + \frac{\omega_{0}}{Q}\dot{x} + \omega_{0}^2x  =  
\frac{\alpha}{m} U+\frac{1}{m}F_{ts}(x). \]

Summarizing the discussion of the tuning fork sensors, we have found 
two reasons why the single harmonic oscillator description is appropriate for 
our sensors: first, the added mass (the glue) is small 
compared to the mass of a single tuning fork arm. Second, the 
relative mechanical coupling strength of the two prongs is strong 
compared to the influence of the added mass and the added spring 
constant. In this respect these 
sensors are completely equivalent to other 
sensors with one prong firmly attached to the support which 
act essentially as an extremely stiff piezoelectric 
cantilever. In other respects there are important differences: we 
find quality factors $Q$ of up to 250000
under the UHV-conditions occuring in our evacuated sample space at a 
temperature of 300 mK. These values are 1-2 orders of magnitude larger
than those reported for the other type of tuning fork sensors 
\cite{giessibl2}. The 
implications of this will be discussed in the remainder of the paper.

\section{Admittance measurement}
\subsection{Model for the admittance}
The mechanical oscillation of tuning fork sensors is measured via the 
piezoelectric effect of the quartz crystal. The induced piezoelectric 
charge on one tuning fork electrode is given by 
$Q_{p}=\alpha\xi=2\alpha x$ and the corresponding piezoelectric 
current is $I_{p}=2\alpha\dot{x}$. In addition, a current 
of size $I_{c}=i\omega CU$ flows through the capacitance between 
the tuning 
fork contacts $C_{0}$ and the total current is therefore given by
$I(\omega) = i\omega C_{0}U+2i\alpha\omega x(\omega)$.
If we neglect the tip-sample interaction force we can determine 
$x(\omega)$ from the harmonic oscillator model and find
\[ I(\omega)= i\omega  U\left(C_{0}+ \frac{2\alpha^2 
/m}{\omega_{0}^2-\omega^2+i\omega\omega_{0}/Q}\right).\]
Resonance curves like the one shown in Fig. \ref{fig67}a) can be 
excellently fitted with this equation. 

Near resonance ($\omega\approx\omega_{0}$) the current is dominated 
by the piezoelectric contribution for our high $Q$ sensors and can be 
expanded to be
\begin{equation}
I(\omega)= \frac{2\alpha^2QU }{m\omega_{0}}
\left(1-i\cdot\frac{2Q}{\omega_{0}}\cdot(\omega-\omega_{0})\right)+{\cal O}\left[(\omega-\omega_{0})^2\right]. 
\label{eq11}
\end{equation}
The component of the current 
which is shifted by 90$^\circ$ with respect to the driving voltage is 
depicted in Fig. \ref{fig67}b). 
It is proportional to the deviation of the excitation frequency from the resonance frequency.
This fact is utilized for the AFM-feedback.

\subsection{Admittance measurement}
We measure this current by using an I-U converter with a current to
voltage conversion ratio of 
$K=10^6$ V/A at 32 kHz. A guard driver neutralizes the huge 
capacitance $C_{K}=1.8$ nF of the long coax-cable connecting the tuning 
fork in the cryostat and thereby 
increases the bandwidth of the I-U converter and reduces the output 
noise. It turns out that the output noise of the converter which was 
found in this setup to be $n_{IUC}=800$ nV/$\sqrt{\mbox{Hz}}$ at 32 kHz,
dominates the noise in the whole AFM feedback.

\begin{figure}[htbp]
  \centering
  \epsfig{file=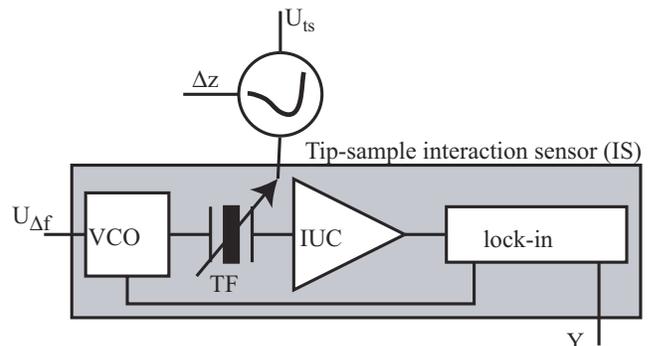,width=3.3in,angle=0}
  \caption{Setup for the measurement of the tuning fork admittance. 
  It acts as the tip-sample interaction sensor (IS) in the SFM.}
  \label{fig2}
\end{figure}

Figure \ref{fig2} shows the setup for the admittance measurement of 
the tuning fork. The fork (TF) is driven by voltage controlled 
oscillator (VCO) with typical excitation amplitudes between 10 $\mu$V and
1 mV depending on the $Q$-value of the tuning fork. The tuning fork current
is converted into a voltage with the I-U converter (IUC). The 
resonance frequency of the tuning fork sensor depends on the tip 
sample interaction via the tip-sample separation $\Delta z$ and the 
tip-sample voltage $U_{ts}$ (see below).
The resulting voltage is demodulated by the lock-in 
amplifier with sensitivity $S$ which determines the in-phase (X) and 
90$^\circ$ (Y) current components corresponding to the real and 
imaginary components in eq. (\ref{eq11}). The output voltages up to 
first order in the frequency shift are 
given by
\begin{eqnarray*}
X_{res} & = & \frac{10\mbox{V}}{S}\cdot K\cdot\frac{\alpha^2QU}{\pi mf_{0}} \\
Y_{res} & = & -\frac{10\mbox{V}}{S}\cdot 
K\cdot\frac{\alpha^2Q^2U}{\pi^2 mf_{0}^2}\cdot\Delta f = 
-X_{res}\cdot\frac{Q}{\pi f_{0}}\cdot\Delta f,
\end{eqnarray*}
where we 
have introduced $f_{0}=\omega_{0}/(2\pi)$ and the frequency shift $\Delta f=f-f_{0}$.
The $Y$-component of the signal is 
a linear indicator of the deviation from the resonance frequency which 
can be utilized for the AFM-feedback.

The spectral noise density on the output signals of the lock-in is 
given by $n_{Y}=10\mbox{V}/S\cdot n_{IUC}$. The integrated output noise
$\delta Y_{res}$ depends on the lock-in time constant $\tau$ and on the order $n$
of its low-pass filter via
\[ \delta Y_{res} = 
c_{n}\times\frac{10\mbox{V}}{s}\times\sqrt{\frac{n_{IUC}^2}{2\pi\tau}}.\]
The constants $c_{n}$ are calculated to be $c_{1}=1.25331$, $c_{2}=0.886227$, 
$c_{3}=0.767495$, $c_{4}=0.700624$, $c_{5}=0.655374$, $c_{6}=0.621742$.
In our setup we use a lock-in time constant of $\tau=100$ $\mu$s corresponding to 
a bandwidth of 1.6 kHz and a fourth order low-pass filter. With 
$S=10$ mV this leads to $n_{Y}=800$ $\mu$V/$\sqrt{\mbox{Hz}}$ and
to typical values of $\delta Y_{res}=22$ mV corresponding to a 
frequency noise of 0.3 mHz.

\subsection{Frequency shift and tip-sample interaction}
The harmonic oscillator approximation for the tuning fork sensors 
allows us to apply the Hamilton-Yacoby perturbation result for 
the frequency shift \cite{giessibl5} 
\begin{eqnarray}
\lefteqn{\Delta f(\Delta z,U_{ts}) =}   \nonumber\\
 & & -\frac{f_{0}}{2\pi Ak}\int_{0}^{2\pi}dx\; F_{ts}\left(\Delta z+A\sin 
x,U_{ts}\right)\sin x. \nonumber\\
 & \approx & \alpha\cdot\Delta U_{ts}+\beta\cdot\Delta z,\label{eq100}
\end{eqnarray}
where $\alpha=\partial\Delta f(0,U_{0})/\partial U_{ts}$ and $\beta=\partial\Delta 
 f(0,U_{ts})/\partial\Delta z$.
The approximate expression of the frequency shift has been obtained by 
an expansion to first order in the tip-sample separation $\Delta z$ and the 
tip-sample voltage $U_{ts}$ which generates electrostatic tip-sample 
forces. The action of these quantities on the admittance measurement 
are indicated in Fig. \ref{fig2}. 

\section{Demodulation of the tuning fork sensor on resonance}
\subsection{Step Response of the Sensor}
For the AFM-operation it is crucial to determine the bandwidth of the 
tip-sample interaction sensor, i.e., the time it takes to respond to a 
given step in $\Delta z$ or $U_{ts}$. We first determine the step 
response of our sensor similar to Albrecht \cite{albrecht}.
Following eq. (\ref{eq1}) we assume that the oscillator is at all times driven by the force
$2\alpha U_{0}\cos(\omega_{e}t+\Phi)$.
For $t<0$ we excite on resonance, i.e. $\omega_{e}=\omega_{0}$ and 
assume that the oscillator performs its steady state oscillations. 
At $t=0$ we instantaneously change the resonance frequency of the 
oscillator to the new frequency $\Omega_{0}=\omega_{0}+\Delta\omega$
and seek solutions of eq. (\ref{eq1}), neglecting $F_{ts}$, of the form
\[ \xi(t) = Y(t)\cos(\omega_{e}t+\Phi) + X(t)\sin(\omega_{e}t+\Phi) \]
For $t<0$ the Amplitudes are 
$X(t)=A_{<}$ and $Y(t)=0$. For $t>0$ we find
\begin{eqnarray}
 Y(t) & = & A_{>}\cos\phi+Be^{-\omega_{0}t/(2Q)}\cos[\Delta\omega t+\theta-\Phi]\nonumber\\
 X(t) & = & -A_{>}\sin\phi-Be^{-\omega_{0}t/(2Q)}\sin[\Delta\omega t+\theta-\Phi]\nonumber
\end{eqnarray}
The constants $B$ and $\theta-\Phi$, found from the boundary 
conditions at $t=0$, are given by
\begin{eqnarray}
B^2 & = & A_{>}^2 + A_{<}^2 + 2A_{>}A_{<}\sin\phi  \nonumber\\
\cot(\theta-\Phi) & = & \frac{A_{>}\cos\phi}{A_{<}+A_{>}\sin\phi}\nonumber
\end{eqnarray}
So far the results are generally valid and no approximations have 
been made. Now we consider approximate behaviour.
For small frequency steps 
$\Delta\omega\ll\omega_{0}/(2Q)$ we find that
\[ Y(t) =  
2A_{<}\frac{Q\Delta\omega}{\omega_{0}}\left[1-e^{-\omega_{0}t/(2Q)}\cos\left(\Delta\omega t+\frac{2Q\Delta\omega}{\omega_{0}}\right)\right]\]
Since the long-time behaviour of the response will be governed by the exponential 
decay we may approximate the response by
\[ Y(t) = 2A_{<}\frac{Q\Delta\omega}{\omega_{0}}\left[1-e^{-\omega_{0}t/(2Q)}\right],\]
i.e., a first order low-pass behaviour with bandwidth $f_{G}=f_{0}/(2 Q)$.
This gives immediately the step response of the demodulated current 
\[ Y_{res}(t) = -X_{res}\cdot\frac{Q}{\pi f_{0}}\cdot\Delta 
f\cdot\left[1-e^{-\pi f_{0}t/Q}\right] ,\]
or, after Fourier transformation
\[ Y_{res}(\omega) =  k(\omega)\Delta f=-X_{res}\cdot\frac{Q}{\pi f_{0}}\cdot\Delta 
f\cdot\frac{1}{1+i\omega/(2\pi f_{G})}.\]

\subsection{Measurement of the Sensor Response}
\begin{figure}[htbp]
  \centering
  \epsfig{file=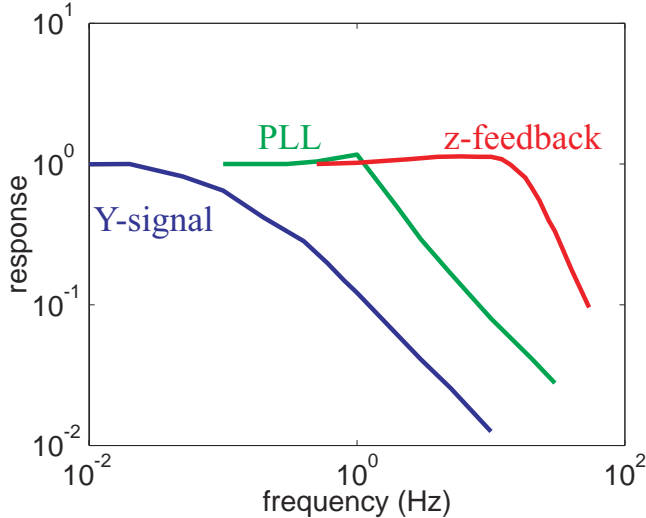,width=3.4in,angle=0}
  \caption{Various response functions of our feedback system. The 
  curve labeled `Y-signal' is the response function of the bare 
  tuning fork sensor demodulated by the lock-in amplifier. The curve 
  labelled `PLL' is the response function of the phase-locked loop. The 
  curve labeled `z-feedback' is the response function of the 
  $z$-feedback including the phase-locked loop.}
  \label{fig100}
\end{figure}

A measurement of this low-pass behaviour is shown in Fig. \ref{fig100}. 
The measurement was performed at a temperature of 4.2 K with the tip at a distance of  80 
nm from the surface of a GaAs/AlGaAs heterostructure in which a 
two-dimensional electron gas (2DEG) resides 34 nm below the surface.
The tuning fork was driven on resonance with an AC-voltage of 
10 $\mu$V applied to the tuning fork contacts. The 
2DEG was employed as the metallic counter electrode of the tip. A DC
tip-sample voltage of $U_{ts}=-9$ V was applied to the 2DEG while keeping 
the tip grounded. Following the idea of eq. (\ref{eq100}) a
low-frequency AC-voltage $\Delta U_{ts}=300$ mV
was added in order to modulate the resonance frequency of the tuning 
fork. Fitting the low-pass behaviour (not shown) results in the characteristic 
frequency of the tuning fork $f_{G}=0.13$ Hz corresponding to 
$Q=116000$. In other experiments at low temperatures we achieved 
$Q$-values of up to 250000.

In order to summarize the results of this section we want to 
emphasize the following points: the $Y$-output of the lock-in 
amplifier is used as the frequency detector in our setup. For small 
changes in resonance frequency $\Delta f\ll f_{G}=f_{0}/(2Q)$ the 
response shows low pass behaviour. Larger changes in resonance 
frequency lead to an oscillatory step response of the phase signal.
High $Q$ sensors have a stronger tendency to an oscillatory response 
due the their higher $f_{G}$ as compared to lower $Q$ sensors.

\section{FM-detection with a phase-locked loop}
FM detection was introduced in dynamic force microscopy by Albrecht 
and co-workers \cite{albrecht}. A theoretical description is due to
D\"urig and co-workers \cite{durig1}. In such setups an FM 
demodulator is used for measuring the oscillation frequency of the 
cantilever. Phase-locked loops are one among many ways of FM demodulation.
They are an established method of frequency measurement 
\cite{tietze,pll}.
The phase-locked loop was first employed in scanning probe 
measurements by D\"urig and co-workers \cite{durig}.  
A digital version of a PLL has been developed by Ch. Loppacher and 
co-workers \cite{lopp}.

Due to the high $Q$-values of our sensors the response to changes in 
resonance frequency, in particular during the approach of the tip to 
the surface, is too slow to allow reasonable approach speeds. This is 
one reason why it is advantageous to employ a phase-locked loop in our 
setup. As we will show below, the phase-locked loop (PLL) increases the 
bandwidth of the response at the expense of increased noise.

\begin{figure}[htbp]
  \centering
  \epsfig{file=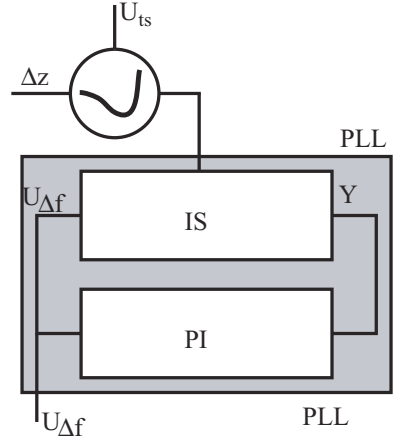,width=2in,angle=0}
  \caption{Phase-locked loop (PLL) comprising the interaction sensor (IS) and the 
  P-I controller (PI).}
  \label{fig3}
\end{figure}

\subsection{Linear response}
The PLL-setup is schematically shown in Fig. \ref{fig3}. The output of the
interaction sensor (IS), i.e., 
the phase signal $Y$, is fed into a P-I controller (PI). Its output 
$V_{\Delta f}$ is used as the input of the VCO in the interaction 
sensor. In the following we describe the PLL by the linear
equations
\begin{eqnarray}
    Y(\omega) & = & k(\omega)(f_{res}(\omega)-f_{d}(\omega))\nonumber\\
    V_{\Delta f}(\omega) & = & R(\omega)Y(\omega)\nonumber\\
    f_{d}(\omega)-f_{0} & = & \eta V_{\Delta f}(\omega),\nonumber
\end{eqnarray}
where $f_{res}$ is the resonance frequency of the tuning fork given by
external parameters such as $\Delta z$ and $U_{ts}$, $f_{d}$ is 
the frequency output of the VCO driving the tuning fork and $f_{0}$ is 
the center frequency of the VCO around which $f_{d}$ can be modulated.
The function $k(\omega)$ is the response of the interaction sensor shown in 
Fig. \ref{fig2} including the lock-in demodulator with its 
frequency response defined by the lock-in time constant and filter 
characteristics, $R(\omega)$ is the response function of the P-I 
controller and $\eta$ is the response function of the VCO which is 
assumed to be frequency independent.
The total response of the PLL depends crucially on the characteristic 
frequency of the P-I controller, $f_{PI}$, its gain $P$ and on the 
characteristic frequency of the tuning fork, $f_{G}$. Parameters are 
optimized, if
\begin{equation}
 f_{PI}=f_{G}.
 \label{eq10}
\end{equation}
If $f_{PI}<f_{G}$ the response of the PLL is unnecessarily retarded 
while for $f_{PI}>f_{G}$ the response function develops an overshoot 
and tends to become unstable.
 If we regard $V_{\Delta f}$ as the output and $f_{res}$ as the input 
of the PLL we find the solution of the linear response equations
\[ V_{f}(\omega) = 
PLL(\omega)\cdot\left(f_{res}(\omega)-f_{0}\right) \]
with the PLL-response function
\[ PLL(\omega) = \frac{1}{\eta}\cdot\frac{L(\omega)}{1+L(\omega)} \]
and the open loop response defined as
\[ L(\omega):=R(\omega)k(\omega)\eta. \]
The response becomes unstable at frequency $\omega_{0}$ if $\mbox{Arg}[L(\omega_{0})]=\pi$ and
$\mbox{Abs}[L(\omega_{0})]>1$ (positive feedback). Stable feedback is 
therefore only achieved if $P$ is kept below a critical value $P_{c}$.
In practice $P_{c}$ has to be determined experimentally.

At very low frequencies $R(\omega)$ is huge and 
$L(\omega)\gg 1$. In this limit $PLL(\omega)$ is 
constant with the value $1/\eta$. At frequencies much larger than 
$f_{PI}=f_{G}$ the function $R(\omega)$ is constant but $k(\omega)$ 
becomes very small such that $L(\omega)\ll 1$ and the response function decays like 
$k(\omega)$.

We now discuss the behaviour of this response function for the case 
that $f_{G}=f_{PI}$. In this case it reduces to
\[ PLL(\omega) = \frac{1}{\eta}\cdot\frac{1}{1+i\cdot\frac{f}{Pa\eta 
f_{G}}}=\frac{1}{\eta}\cdot\frac{1}{1+i\cdot\frac{f}{f_{PLL}}}, \]
where we have introduced $a=X_{res}Q/(\pi f_{0})$,
i.e. to a low-pass behaviour with the new characteristic frequency
\[ f_{PLL} = P a \eta f_{G}. \]
The product $a\eta$ acts like an additional $P$-factor and the 
effective $P$-factor is $P_{eff}=Pa\eta$. The bandwidth of the PLL is 
a factor of $P_{eff}$ larger than the bandwidth of the interaction 
sensor itself. This is shown in Fig. \ref{fig100} as the curve 
labeled `PLL' which was measured with $\eta=0.1$ Hz/V, $a=71$ V/Hz 
and $P=2$. Higher bandwidths can be achieved with higher $P_{eff}$, 
however, the stability condition $P_{eff}<P_{c}$ and noise 
considerations constitute the upper experimental limits.

\subsection{Noise considerations}
The dominating noise from the I-U converter can be represented as an 
equivalent noise source $n_{Y}(\omega)$ at the Y-output of the lockin.
A linear response analysis starts from the equations
\begin{eqnarray}
    Y(\omega) & = & 
    k(\omega)(f_{res}(\omega)-f_{d}(\omega))+n_{Y}(\omega)\nonumber\\
    V_{f}(\omega) & = & R(\omega)Y(\omega)\nonumber\\
    f_{d}(\omega)-f_{0} & = & \eta V_{f}(\omega).\nonumber
\end{eqnarray}
The resulting noise term of the solution for $f_{G}=f_{PI}$ is
\begin{equation}
n_{PLL}(\omega) = P\cdot\frac{f-i\cdot f_{G}}{f-i\cdot f_{PLL}}\cdot 
n_{Y}(\omega).
\label{npll}
\end{equation}
At frequencies $f\gg f_{PLL},f_{G}$ 
the noise response is simply given by $P$, while for $f\ll 
f_{PLL},f_{G}$ the noise is given by the ratio 
$f_{G}/f_{PLL}=1/(Pa\eta)<1$. The low frequency noise is reduced by 
the PLL. However, since $f_{PLL}$ is small compared to the lockin 
bandwidth of about 530 Hz, the low frequency noise suppression is not 
relevant in practice for the integrated noise spectrum and to a good 
approximation the output noise of the PLL $\delta V_{\Delta f}$ is enhanced by the factor $P$
over the noise on the $Y$-output $\delta Y$:
\begin{equation}
\delta U_{\Delta f} = P\cdot \delta Y.
\label{eq33}
\end{equation}

\begin{figure}[htbp]
  \centering
  \epsfig{file=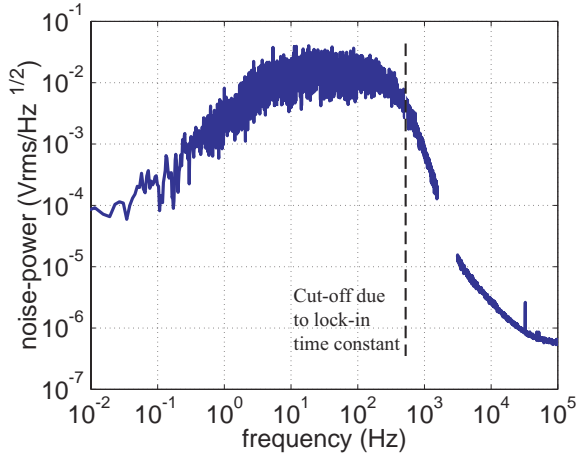,width=3 in,angle=0}
  \caption{Output noise of the PLL in the full range of 
  frequencies}\label{fig15}
\end{figure}

A measurement of the PLL-noise is shown in Fig. \ref{fig15}. 
Here, $\eta= 0.1$ Hz/V, the tuning fork excitation was 10 $\mu$V.
The lockin time constant $\tau=300$ $\mu$s at 24 dB/oct is 
responsible for the strong cut-off of the noise spectrum above 500 Hz.
The I-U converter output noise $n_{IUC}$ was amplified by a factor 1000 by the 
lock-in and therefore has a value of 800 $\mu$V/$\sqrt{\mbox{Hz}}$.
In the frequency range between 5 and 500 Hz the PLL-noise is enhanced 
by the factor $P=12$. At the lowest frequencies the noise is reduced 
in agreement with the above discussion. The integrated noise spectrum 
corresponds to an effective frequency noise of about 20 
mHz/$\sqrt{\mbox{Hz}}$.
If desired, parameters can be chosen such that the frequency noise is 
suppressed well below 1 mHz/$\sqrt{\mbox{Hz}}$ at the cost of bandwidth. For 
scanning we typically work with a PLL-noise of 10 mHz/$\sqrt{\mbox{Hz}}$.

\section{The $z$-feedback}
\subsection{Linear response}
\begin{figure}[htbp]
  \centering
  \epsfig{file=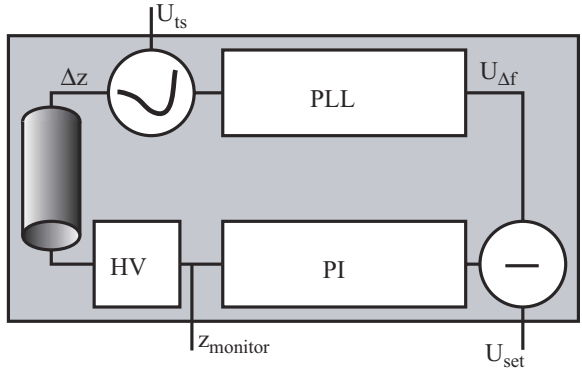,width=3 in,angle=0}
  \caption{Scheme of the $z$-feedback consisting of the phase-locked loop (PLL), 
  the P-I controller (PI), the high-voltage amplifier (HV) and the 
  scan piezo tube.}
  \label{fig4}
\end{figure}

The PLL-feedback is part of the $z$-feedback loop as shown in Fig. 
\ref{fig4}. The PLL-output voltage $U_{\Delta f}$ is fed into the 
$z$-feedback P-I controller (PI) which drives the high-voltage 
amplifier (HV). This  amplifier supplies the voltage for the $z$-piezo electrodes
and thereby determines the tip-sample separation $\Delta z$. As a 
result of the 
force-distance interaction characteristic, $\Delta z$ and the 
tip-sample voltage $U_{ts}$ are translated into a resonance frequency of the 
interaction sensor [see eq. (\ref{eq100})] which is part of the PLL.
For controlling the tip at constant height above the sample surface a 
certain $U_{set}$ is chosen corresponding to a certain frequency 
shift via $f_{set}=\eta U_{set}$. The feedback will then keep the 
tuning fork resonance frequency at $f_{set}$ by controlling $\Delta z$ 
during a scan.

In the following we discuss the $z$-feedback in a similar fashion to 
the PLL. Here the relevant linear equations are
\begin{eqnarray}
    \Delta f(\omega) & = & \alpha U_{ts}^{AC}(\omega)+\beta\Delta 
    z(\omega) \nonumber\\
    \Delta z(\omega) & = & \mu\cdot \Delta U_{HV}(\omega) \nonumber\\
    U_{\Delta f}(\omega) & = & PLL(\omega)\cdot\Delta f(\omega) \nonumber\\
    e(\omega) & = & U_{set}-U_{\Delta f}(\omega) \nonumber\\
    \Delta U_{HV}(\omega) & = & PI(\omega)\cdot e(\omega) \nonumber
\end{eqnarray}
Here $\mu$ is the (temperature-dependent) calibration of the 
$z$-piezo tube (e.g. 425 nm/70 V at 4.2 K), $U_{set}$ is the setpoint 
of the feedback, $PI(\omega)$ is the response of the $z$-feedback
P-I controller and $e(\omega)$ is the error signal of the feedback 
loop. We are interested in the response 
of $\Delta z$ as a result of surface roughness on the sample. 
We can simulate this roughness by applying a sinusoidal AC tip-sample 
voltage $V_{ts}^{AC}$. The solution is
\[ \Delta z = Z(\omega)\cdot\frac{\alpha}{\beta}\cdot U_{ts}^{AC}. \]
Here, the quantity $\alpha/\beta\cdot U_{ts}^{AC}$ plays the role of 
an effective surface roughness $\Delta z_{eff}$ and
\[ Z(\omega) =  -\frac{M(\omega)}{1+M(\omega)}.\]
with the open loop response
\[ M(\omega) := \beta\mu PI(\omega)PLL(\omega). \]

In the limit of low frequencies $PLL(\omega)$ is constant, 
$PI(\omega)$ dominates and $\beta\mu PI(\omega)PLL(\omega)\gg 1$ such 
that $Z(\omega)$ is constant and given by $-z_{eff}$.
In the limit of large frequencies this response 
decays like $PLL(\omega)$.

This feedback is stable, if $\mbox{Abs}[M(\omega)]<1$ at the 
frequency where $\mbox{Arg}[M(\omega)]=\pi$. This gives again an 
upper limit $P_{c}^{(z)}$ for possible $P_{z}$ values. $P_{c}^{(z)}$ 
depends on $\beta$ and therefore on the tip-sample interaction.

If we assume again that
\begin{equation}
	f_{PLL}=f_{z}
	\label{eq23}
\end{equation}
the z-feedback response 
simplifies to
\[ Z(\omega) = -\frac{1}{1+i\cdot f/f_{FB}}\]
with
\begin{equation}
f_{FB}=\frac{P_{z}\beta\mu}{\eta}\cdot f_{PLL}=P_{z}P\beta\mu a\cdot 
f_{G}. 
\label{eq13}
\end{equation}
Here the product $\beta\mu a$ acts as an additional $P$-factor and the 
effective value is $P_{eff}^{(z)}=P_{z}P\beta\mu a$ such that
$ f_{FB}=P_{eff}^{(z)}f_{G}$, i.e., the bandwidth of the $z$-feedback 
is again larger than the bandwidth of the PLL. This is demonstrated in 
Fig. \ref{fig100} where the curve labeled `z-feedback' was measured 
by varying the frequency of $U_{ts}^{AC}(\omega)$. The figure 
summarizes the resulting frequency characteristics of the two nested 
feedback-loops, i.e., the PLL and the $z$-feedback. It is demonstrated 
that the total system response is a low-pass, if all parameters 
are properly set. The bandwidth of the $z$-feedback can be increased 
by using higher $P_{eff}$, however, the stability criterion 
$P_{eff}<P_{c}^{(z)}$ and noise considerations constitute the 
experimental limits.

\subsection{Noise considerations}
If we consider the contribution of the dominant I-U-converter noise in 
the $z$-feedback the relevant equations are
\begin{eqnarray}
    \Delta f(\omega) & = & \alpha U_{ts}^{AC}(\omega)+\beta\Delta 
    z(\omega) \nonumber\\
    \Delta z(\omega) & = & \mu\cdot \Delta U_{HV}(\omega) \nonumber\\
    U_{\Delta f}(\omega) & = & PLL(\omega)\cdot\Delta f(\omega) 
    +n_{PLL}(\omega)\nonumber\\
    e(\omega) & = & U_{set}-U_{\Delta f}(\omega) \nonumber\\
    \Delta U_{HV}(\omega) & = & PI(\omega)\cdot e(\omega) \nonumber
\end{eqnarray}
The noise contribution $n_{PLL}(\omega)$ is the PLL 
output noise calculated earlier (see eq. (\ref{npll})) which is dominated by the ouput noise 
of the I-U converter.
The response to the PLL-output noise determined from these equations 
under the condition $f_{PLL}=f_{z}$ is
\[ n_{z}(\omega)=-\mu P_{z}\cdot P\cdot 
\frac{f-if_{G}}{f-if_{0}}\cdot n_{Y}(\omega). \]
It is interesting to note that the bandwidth of the PLL does not 
enter the noise figure of the $z$-feedback.
The integrated noise of the $z$-feedback is given by
\begin{equation}
\delta z = \mu P_{z}P\cdot\delta Y 
\label{eq101}
\end{equation}
Higher bandwidth of the $z$-feedback will lead to larger $z$-noise 
and for high quality imaging it is important to keep $\delta z$ 
as low as possible (typically of the order of 1 \AA). On the other 
hand, reasonable scan speeds require high bandwidths of the order of 
several hundred Hz. This naturally leads to the question, how the 
optimum feedback parameters can be found from the above analysis
for a given setup.

\section{Optimum feedback parameters}

The optimum feedback parameters are given by the following conditions:
\begin{enumerate}
    \item The characteristic frequency of the PLL-PI-controller is 
    identical to the characteristic frequency $f_{G}$ of the (tuning 
    fork) sensor.
    \item The characteristic frequency of the $z$-feedback P-I 
    controller is identical to the PLL-bandwidth. 
    \item The $P_{eff}$ parameter of the PLL is well below the critical 
    value $P_{c}$ where the PLL-feedback becomes unstable.
    \item The $P_{eff}^{(z)}$ parameter of the z-feedback is well below the critical 
    value $P_{c}^{(z)}$ where the z-feedback becomes unstable.
    \item The $P_{eff}$ parameter of the PLL is small enough to give 
    acceptable frequency noise.
    \item The $P_{eff}^{(z)}$ parameter of the $z$-feedback is small 
    enough to give acceptable $z$-noise.
\end{enumerate}
We repeat the corresponding equations (\ref{eq10}), (\ref{eq23}),  
(\ref{eq33}) and (\ref{eq101}) below:
\begin{eqnarray}
 f_{PI} & = & f_{G}=\frac{f_{0}}{2Q} \label{r1}\\
 f_{z}  & = & f_{PLL}=P\eta\frac{X_{res}}{2\pi} \label{r2} \\
 \delta U_{\Delta f} & = & P\cdot\delta Y \label{r3}\\
 \delta z & = & \mu P_{z}P \cdot\delta Y \label{r4}
 \end{eqnarray}
 From these equations the optimum feedback parameters can be found as 
 follows: First, the tuning fork resonance curve is measured and $f_{G}$ 
 is determined from its half width at the $\sqrt{2}$-maximum value. 
 This frequency determines the characteristic frequency to be set on 
 the P-I controller of the PLL via eq. (\ref{r1}). The gain $P$ of this 
 P-I controller is set such that the output noise of the PLL 
 [$\delta U_{\Delta f}$, eq. (\ref{r3})] corresponds to a reasonable frequency 
 noise (typically 10 mHz for scanning in our setup). Next, the P-I 
 parameters of the $z$-feedback are set such that eq. (\ref{r2}) is 
 fulfilled. Finally, the gain $P_{z}$ of the $z$-feedback is adjusted 
 according to the tolerable $z$-noise with the help of eq. 
 (\ref{r4}). We typically keep the $z$-noise around 1 \AA\ for normal 
 scanning operation.
 
 We emphasize that the optimum 
 feedback parameters determined in this way do not depend on the tip 
 shape or any details of the tip-sample interaction. Knowledge of the 
 resonance curve of the sensor alone is sufficient for the optimum 
 settings of the feedback.
 
\begin{figure}[htbp]
  \centering
  \epsfig{file=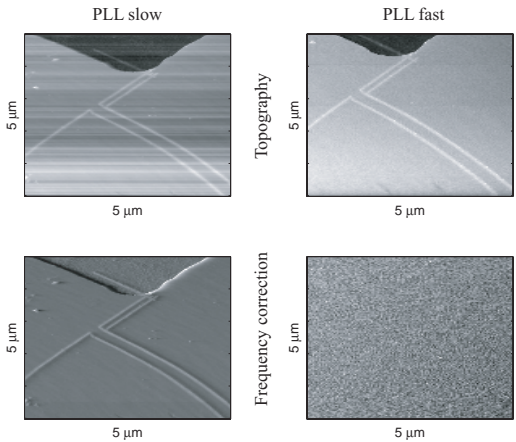,width=3.4 in,angle=0}
  \caption{Left column: Topographic image (top) and frequency 
  error signal (bottom) with small PLL bandwidth. Right column: the 
  same with large PLL bandwidth.}
  \label{fig49}
\end{figure}

 For a given $z$-noise there is a certain freedom concerning the 
 choice of the parameters $P$ and $P_{z}$, since according to eqs. 
 (\ref{r3}) and (\ref{r4}) only their product needs to be kept constant.
 This freedom allows us to choose the bandwidth of the PLL either 
 relatively high or relatively low in the span between $f_{G}$ and 
 $f_{FB}$. If the PLL-bandwidth is chosen to be high, e.g. close to 
 $f_{FB}$, the PLL will be able to track the resonance frequency with 
 the same bandwidth as the $z$-feedback and $U_{\Delta f}$ will be 
 constant throughout a scan. However, if the PLL bandwidth is chosen 
 to be low, e.g. closer to $f_{G}$, the PLL can not track the 
 resonance frequency with the same bandwidth as the $z$-feedback and 
 appreciable error signals arise whenever there is a step on the surface.
 This is demonstrated in Fig. \ref{fig49}. The two images in the top 
 row are topographic images of a detail of a Hall bar structure with 
 AFM-written oxide lines in the bottom right corner. The left (right) image 
 was taken with small (large) PLL-bandwidth at fixed bandwidth of 
 the $z$-feedback. The bottom row shows the corresponding frequency 
 error signal. It can be seen that with a slow PLL, an additional 
 image can be obtained in which the contours of edges appear very sharp. 
 With a fast PLL this contrast disappears completely. This 
 `differential' image of the surface can be very useful since it 
 eliminates sample tilt and amplifies edges of small height nicely.

\section{Effect of Q and k on the bandwidth}

It becomes clear from the above discussion that
the bandwidth of the $z$-feedback is given by eq. (\ref{eq13})
\[ f_{FB}=P_{z}P\beta\mu a f_{G} = P_{z}P\beta\mu \frac{X_{res}}{2\pi}\]
i.e., it is independent of the $Q$-factor of the tuning fork cantilever.
This is not obvious, since high $Q$ cantilevers have a low $f_{G}$ and 
are therefore known to slow down the response. However, implemented in 
our nested feedback system the quality factor of the sensor is 
canceled from the final expression for the bandwidth. The only impact 
on the performance stems from the fact that the sensor response can 
only be described as a low-pass, if changes in frequency $\Delta f$ typically 
occurring during a scan are small compared to $f_{G}=f_{0}/(2Q)$.

The situation is different with the stiffness $k$ of the sensor which 
directly enters the quantity $\beta$, i.e., the slope of the $\Delta 
f(z)$ curve at the operating point. Softer sensors will lead to larger 
values of $\beta$ and thereby increase the bandwidth of the 
$z$-feedback.

Similar arguments play a role for the comparison of scanning in the 
attractive or repulsive part of the $\Delta f(z)$ curve. In the 
attractive part typical values of $\beta$ are much smaller than in the 
repulsive part. Therefore the bandwidth tends to be considerably larger
for controlling in the repulsive regime.

\section{Comparison to phase control}

We briefly want to discuss why the use of the PLL is advantageous 
compared to phase control, where the phase change measured as the 
$Y$-output of the lock-in in our case is directly used for the 
$z$-controller without the intermediate PLL.

From a similar linear response analysis as performed in this paper for 
our nested system one finds that the same bandwidths of the 
$z$-feedback can be achieved with pure phase control. Enhancement of 
the bandwidth for scanning operation is therefore not the striking 
argument for the use of the additional PLL feedback.

However, there are several reasons, why we use the PLL:
\begin{enumerate}
	\item The PLL increases the bandwidth of the tuning fork response 
	when the $z$-feedback is not yet controlling, e.g. during the 
	approach of the tip. The PLL allows the use of reasonable approach 
	speeds.
	\item The PLL increases the dynamic range of the frequency 
	detection. The Y-signal of the lock-in depends linearly on the 
	frequency shift only in a limited range of frequencies around 
	resonance. With high $Q$ sensors this range can become so narrow 
	(100 mHz) that during tip approach the frequency shift runs out 
	of this range. The PLL avoids this problem by tracking the resonance. 
	The frequency range that can be used with PLL is mainly determined by 
	the VCO-settings. We typically use a range of 2 Hz.
	\item The PLL allows to image with an additional adjustable frequency 
	contrast (see Fig. \ref{fig49}).
	\item When one measures $\Delta f(z)$-curves the PLL allows the 
	clear distinction between conservative frequency shifts and 
	dissipative effects \cite{durig1,rychen3}.
\end{enumerate}

\section{Conclusion}
In the first part of this paper we have shown that the tuning fork 
sensors that we use in our cryo-SFM setups can be well modeled by 
a single-harmonic oscillator despite the symmetry breaking wire that 
is attached to one tuning fork prong. In the second part of the paper 
we have analyzed the linear response of our nested feedback system 
comprising the interaction sensor, a phase-locked loop and a 
conventional $z$-feedback. We have demonstrated that the optimum 
feedback parameters of such a setup can be easily found and that they 
are independent of the tip shape and of the details of the tip-sample 
interaction. We have compared our setup to the conventional 
phase-control mode and discussed the advantages of the use of the 
phase-locked loop for high-Q cantilevers.

\end{multicols}
\end{document}